%% file: paper.tex
\documentclass[sigconf]{acmart}

\usepackage{booktabs} 

\renewcommand\footnotetextcopyrightpermission[1]{} 
\usepackage{mathtools}

\usepackage{arydshln,leftidx}
\usepackage{amsmath}
\usepackage{amssymb}
\usepackage{kbordermatrix}
\usepackage{graphics}
\usepackage{graphicx}
\usepackage{enumitem}
\usepackage{textcomp}
\usepackage{color}
\usepackage{outlines}
\usepackage{listings}
\usepackage[normalem]{ulem}

\usepackage{amsthm}

\usepackage{caption}
\usepackage{setspace}
\usepackage{threeparttable}

\usepackage{changepage}

\usepackage{etoolbox}
\let\bbordermatrix\bordermatrix
\patchcmd{\bbordermatrix}{8.75}{4.75}{}{}
\patchcmd{\bbordermatrix}{\left(}{\left|}{}{}
    \patchcmd{\bbordermatrix}{\right)}{\right|}{}{}

\usepackage{blkarray}
\usepackage{multirow}
\usepackage{wrapfig}
\usepackage{tikz}

\usepackage{subfig}
\usetikzlibrary{matrix,fit}

\begin{document}
\title{CodeTrolley: Hardware-Assisted Control Flow Obfuscation\vspace{-1.0in}}
 \author{Novak Boskov, Mihailo Isakov and Michel A. Kinsy\\
 Adaptive and Secure Computing Systems (ASCS) Laboratory\\
 Department of Electrical and Computer Engineering\\
Boston University\\
\{boskov, mihailo, mkinsy\}@bu.edu}

\begin{abstract}
  Many cybersecurity attacks rely on analyzing a
  binary executable to find exploitable sections of code.
  Code obfuscation is used to prevent attackers from reverse
  engineering these executables.
  In this work, we focus on control flow obfuscation - a technique
  that prevents attackers from statically determining which code
  segments are original, and which segments are added in to confuse
  attackers.
  We propose a RISC-V-based hardware-assisted deobfuscation technique
  that deobfuscates code at runtime based on a secret safely stored in hardware,
  along with an LLVM compiler extension for obfuscating binaries.
  Unlike conventional tools, our work does not rely on compiling hard-to-reverse-engineer
  code, but on securing a secret key. As such, it can be seen as a lightweight alternative
  to on-the-fly binary decryption.
  \vspace{-0.1in}
\end{abstract}


\maketitle

\input{intro}
\input{obfuscate}

\input{deobfuscate}
\input{evaluate}
\input{concl}

\bibliographystyle{ACM-Reference-Format}
\bibliography{paper}

\end{document}

%% file: intro.tex
\vspace{-0.1in}
\section{Introduction}
Attackers may want to steal sensitive data or cryptographic keys from a
system. To mount their attacks they have to find an appropriate
location in the targeted software. 
Attackers attempt to reverse engineer the targeted executables in order to find exploitable sections of code.
To prevent reverse engineering, software vendors resort to various kinds of software obfuscations. One
of the most effective obfuscation targets is the program's control flow (CF). 
CF obfuscation is a technique in which a dedicated program called the obfuscator performs semantic-preserving transformations on
the original program in order to hide the original control flow.
This kind of obfuscation heavily relies on opaque predicates. 
A predicate is opaque if its resolution is hard or ambiguous for the attacker. 
The technique of opaque predicates is used in obuscation tools such as
Obfuscator-LLVM~\cite{Junod_2015}. 
The construction of an opaque predicate is usually done by tailoring a computationally 
intensive or even uncomputable challenge for the underlying concolic execution
engine such as present in BAP~\cite{BAP}. 
Some of the challenges proposed
by Xu et. al.~\cite{Xu_2018} are: symbolic memory, floating-point
algebra, covert symbolic propagation and parallel programming. Even
though one may construct multiple different concrete challenges from
the same basic ideas, they rarely pose a theoretical barrier to
reverse engineering but rather technical difficulties. Such technical
difficulties get eliminated by enhancements in concolic
engines as soon as their authors implement the missing part that
allows for the difficulty to surface. Whole-executable encryption is a plausible alternative to control flow
obfuscation. However, this approach incurs its own hazards. The
decryption keys need to be stored securely, and the decrypted binary
cannot be stored in off-chip memory at any moment in time. If an
executable has to be decrypted before execution, that has to be done
in a safe environment (enclave) such as offered by Intel's
SGX~\cite{SGX}. Furthermore, on-the-fly decryption incurs a significant
performance penalty which this work aims to avoid. 
This work proposes the idea of hardware-assisted CF obfuscation whose
integrity relies on a secret that is available only to the trusted
party, like an unclonable hardware module. This approach does not depend on the ability of an attacker or
concolic execution engine to evaluate particular portion of
computation as long as the secret is kept safe.

%% file: obfuscate.tex
\section{Obfuscation Process}
The essence of our approach relies on potentially reversing all the
conditional branches in the original program. The obfuscator decides
whether to revert a branch by calculating a function that takes two inputs:
(1) a ID of the branch, and (2) a program key. If the function returns a 1, 
the branch condition is reversed, otherwise it is not.
In our implementation, we use a cryptographic hash with a binary output. 
Hence, the only part of secret information is the program key. 

Listing~\ref{lst:ccode} shows a segment of a C program that is
compiled using our branch obfuscator. 
Listing~\ref{lst:plain} shows the assembly of the original program. 
Assuming that the hash function reverses both branches, 
obfuscated RISC-V assembly is shown in Listing~\ref{lst:obf}. 

\begin{tabular}{c}
\begin{lstlisting}[caption=C code of the original program, language=C,
    basicstyle=\smaller, label={lst:ccode}]
    ...
    if (n < 5)
        printf("Your number is lower than 5\n");
    if (n > 12)
        printf("Higher than 12\n");
    ...
\end{lstlisting}
\end{tabular}

\noindent\begin{minipage}{.23\textwidth}
  \begin{tabular}{c}
\begin{lstlisting}[caption=Plain code, basicstyle=\tiny, label={lst:plain}]
      main:
      ...
      addi a1, zero, 4
      blt  a1, a0, .LBB0_2
      j    .LBB0_1
      .LBB0_1:
      lui  a0, %hi(.L.str.2)
      addi a0, a0, %lo(.L.str.2)
      call printf
      j    .LBB0_2
      .LBB0_2:
      lw   a0, -16(s0)
      addi a1, zero, 13
      blt  a0, a1, .LBB0_4
      j    .LBB0_3
\end{lstlisting}
  \end{tabular}
\end{minipage}\hfill
\begin{minipage}{.23\textwidth}
  \begin{tabular}{c}
\begin{lstlisting}[caption=Obfuscated code, basicstyle=\tiny, label={lst:obf}]
      main:
      ...
      <@ \textcolor{red}{addi a1, zero, 5} @>
      <@ \textcolor{red}{blt  a0, a1, .LBB0\_2} @>
      j   .LBB0_1
      .LBB0_1:
      lui  a0, %hi(.L.str.2)
      addi a0, a0, %lo(.L.str.2)
      call printf
      j    .LBB0_2
      .LBB0_2:
      lw   a0, -16(s0)
      <@ \textcolor{red}{addi a1, zero, 12} @>
      <@ \textcolor{red}{blt  a1, a0, .LBB0\_4} @>
      j    .LBB0_3
\end{lstlisting}
  \end{tabular}
\end{minipage}
\vspace{0.1in}

\begin{figure}[h!]
  \vspace{-0.15in}
  \centering
  \includegraphics[width=0.42\textwidth]{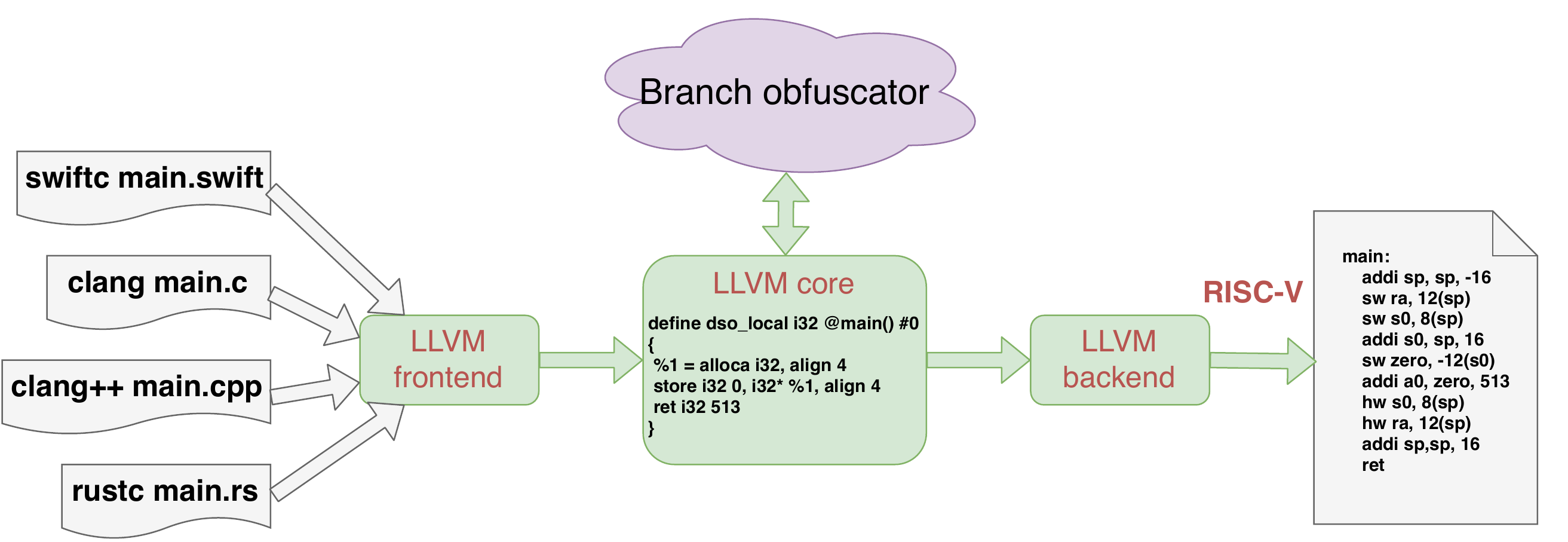}
  \caption{Place of the obfuscator in LLVM compilation chain.}
  \label{fig:obfuscator}
  \vspace{-0.25in}
\end{figure}

The obfuscator itself is implemented as an LLVM Pass
- the same technique that the LLVM compiler infrastructure uses
internally for its optimizations. The place of the obfuscator in the
LLVM compilation chain is given in Figure~\ref{fig:obfuscator}. It
performs branch obfuscation on LLVM internal representation (IR) and thus supports all 
programming languages with compilers that target LLVM IR. 
This group of programming languages includes C, C++, Rust, Apple
Swift and others. The resulting LLVM IR is then passed to the RISC-V
compiler backend that emits obfuscated RISC-V assembly.

%% file: deobfuscate.tex
\section{Deobfuscation Process} \label{lab:deobf}

The obfuscated program correctly executes only on a trusted RISC-V
core designed to support deobfuscation. We outline four designs here:
the baseline, stalled-hash, mask-based, and the cached-hash design.

\begin{figure}[h!]
  \vspace{-0.1in}
  \centering
  \includegraphics[width=0.42\textwidth]{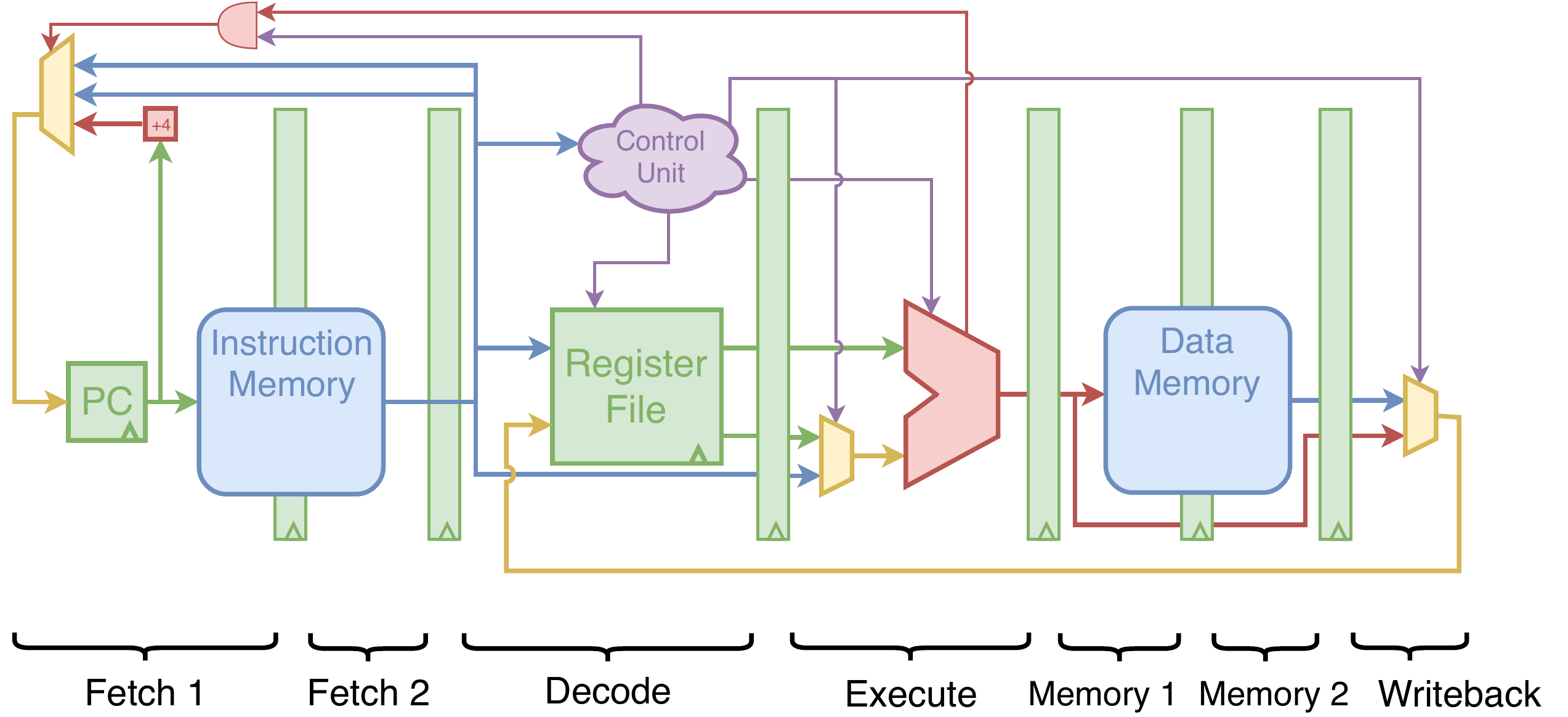}
  \caption{The original 7-stage pipeline RISC-V processor.}
  \label{fig:arch1}
  \vspace{-0.15in}
\end{figure}

\noindent\textbf{Baseline design:} the baseline design is a 7-stage
RISC-V CPU without any hardware modifications enabling obfuscation. A
simplified processor architecture is illustrated in
Figure~\ref{fig:arch1}.

\begin{figure}[h!]
\vspace{-0.1in}
  \centering
  \includegraphics[width=0.42\textwidth]{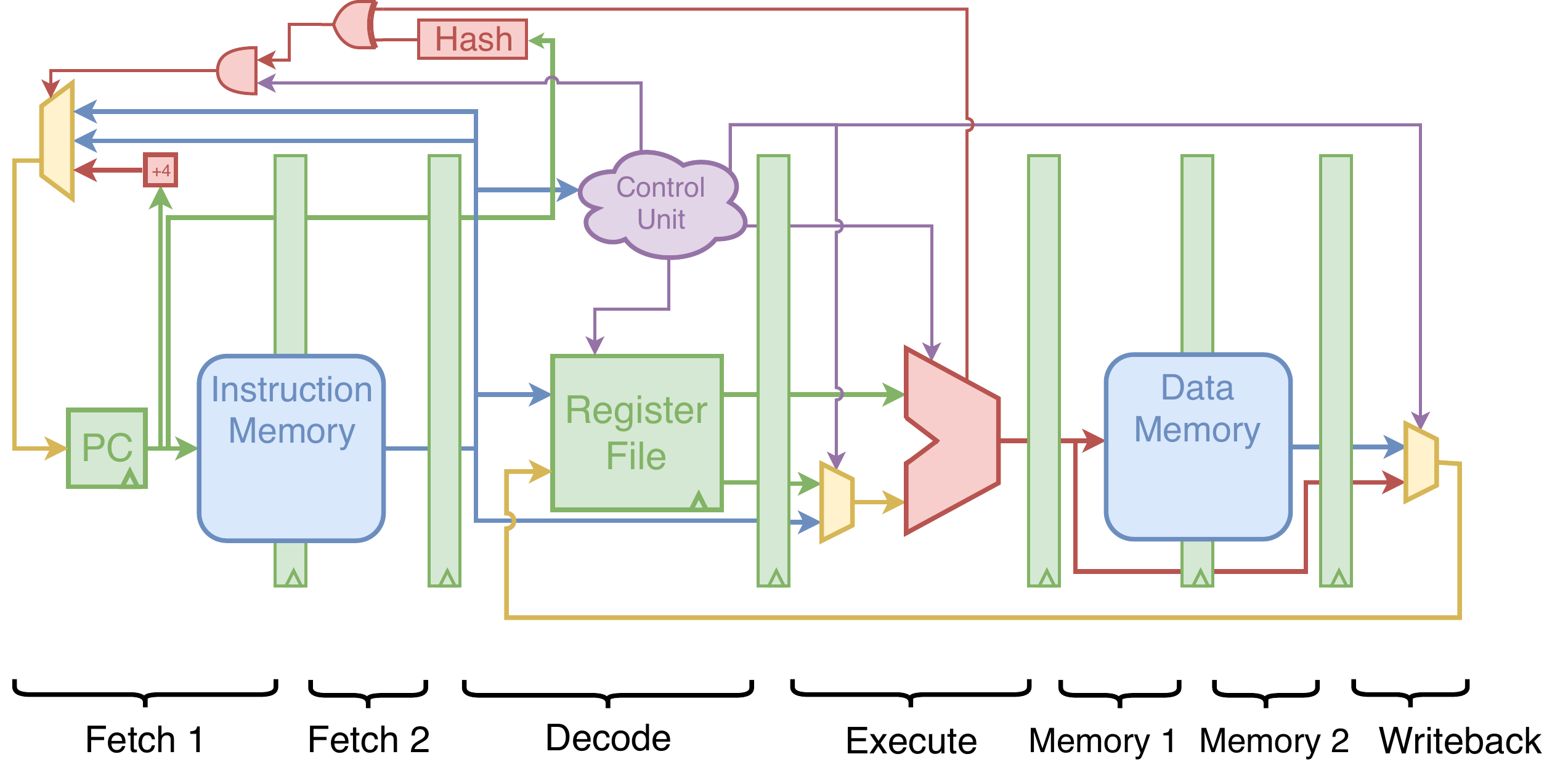}
  \caption{The original 7-stage pipeline RISC-V processor modified to
    XOR branch values with hash function outputs.}
  \label{fig:arch2}
  \vspace{-0.15in}
\end{figure}

\noindent\textbf{Stalled-hash design:} here, the baseline CPU is
equpped with a hardware hash function. When a branch instruction is in
the decode stage, the hash function is fed the branch instruction
address and the program key. When the branch instruction reaches the
execute stage, all the stages up to and including the execute stage
are stalled until the hash function produces an output. Once a (single
bit) output is produced, that value is XOR-ed with the branch
signal. This way, branches that would be taken may not be, and
vice-versa.
The modified architecture is illustrated in Figure~\ref{fig:arch2}.

\noindent\textbf{Mask-based design:}
instructions are extended with a mask bit. The mask bit specifies
whether a branch should be reversed or not.  Having an independent
mask bit per each branch removes the possibility of an attacker
predicting future branches based on existing ones. However, widening
the instruction word width complicates the design of L2 caches and
memory controllers.  Furthermore, the masks need to be kept encrypted
in memory and decrypted on-the-fly.

\noindent\textbf{Cached-hash design:} in this design, we add a
hash-cache to the stalled-hash design. When a branch is in the decode,
the architecture in parallel starts calculating the hash function and
checks the cache for whether that branch's hash has previously been
calculated. If not, when the hash function finishes, it both feeds the
value to the XOR gate, and saves the result in the cache. If the value
is found in the cache, it is sent to the XOR, just in time as the
branch enters the execute stage. In our experiments we used a simple
256-line, one branch per line, direct-mapped cache.

%% file: evaluate.tex
\section{Performance Evaluation}
\begin{figure}[h!]
  \vspace{-0.1in}
  \centering
  \includegraphics[width=\columnwidth]{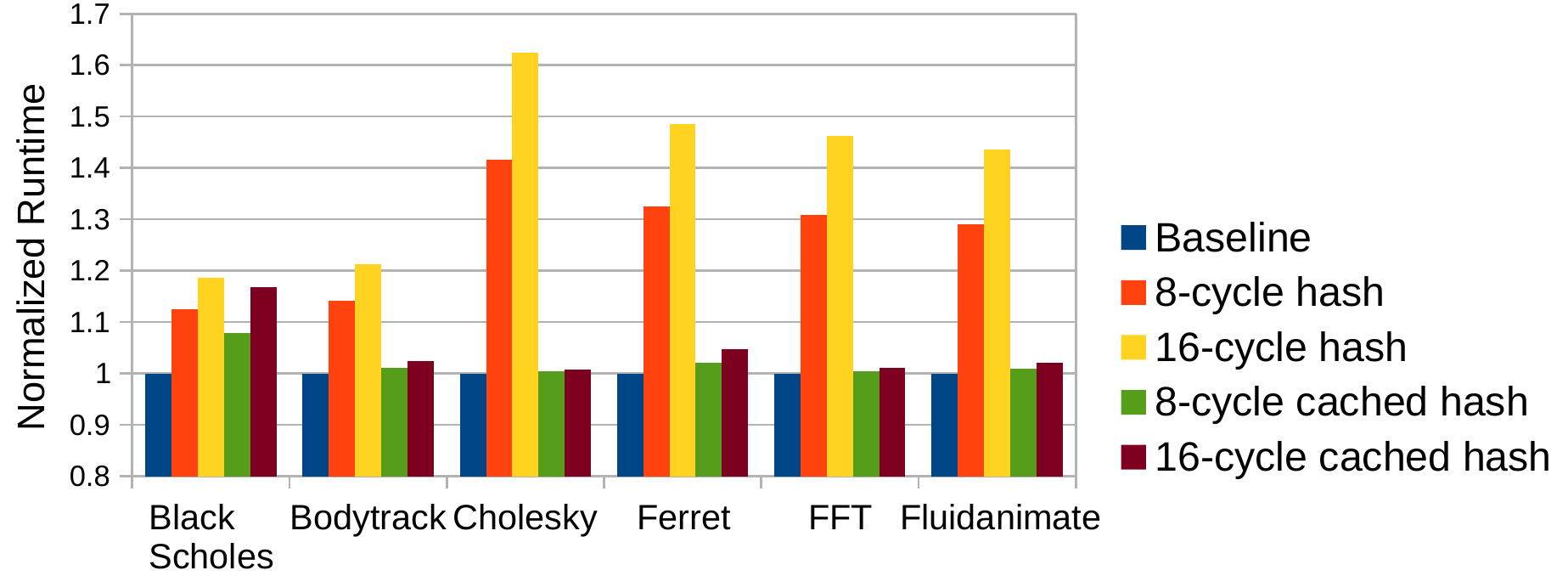}
  \caption{The performance of the five architectures on 6 PARSEC tasks, measured in cycles.}
  \label{fig:eval}
  \vspace{-0.25in}
\end{figure}

We simulate the four architectures listed above. The
baseline and the mask-based architectures have the same performance,
so we omit the second one. Further, we test two different hash-based
architectures with an 8 and a 16-cycle hash function. In
Figure~\ref{fig:eval}, we see the performance of different
architectures on 6 different PARSEC tasks. Notice that for the
16-cycle hash, the processor can slow down as much as 60\%. However,
adding a (256-line, single branch hash per line) direct-mapped cache
removes the majority of the performance overhead.

%% file: concl.tex
\section{Conclusion}
In this work we have explored custom hardware for efficient 
binary control flow obfuscation. We have presented our extensions to
the LLVM compiler which allow simple compile-time obfuscation. Next, we
have shown several architectures that allow on-chip deobfuscation of
code. Finally, we have measured the performance penalty of these
hardware modifications and have shown that our cached-hash-based 
implementation achieves full control flow
obfuscation with a small performance penalty.